\documentclass[11pt]{article}
\usepackage{amsmath,amssymb,color,graphics,epsfig}
\usepackage{graphicx}
\usepackage{subfigure}

\usepackage{amsfonts}

\newcommand{\be}{\begin{equation}}
\newcommand{\ee}{\end{equation}}
\newcommand{\bea}{\setlength\arraycolsep{2pt} \begin{eqnarray}}
\newcommand{\eea}{\end{eqnarray}}
\newcommand{\nn}{\nonumber}
\newcommand{\mc}{\mathcal}
\newcommand{\mm}{\mathrm}

\def\fft#1#2{{\frac{#1}{#2}}}

\def\0{{\sst{(0)}}}
\def\1{{\sst{(1)}}}
\def\2{{\sst{(2)}}}
\def\3{{\sst{(3)}}}
\def\4{{\sst{(4)}}}
\def\5{{\sst{(5)}}}
\def\6{{\sst{(6)}}}
\def\7{{\sst{(7)}}}
\def\8{{\sst{(8)}}}
\def\sst#1{{\scriptscriptstyle #1}}

\begin{document}

\begin{flushright}
\end{flushright}

\vspace{25pt}
\begin{center}
{\large {\bf  The growth of operator entropy in operator growth}}

\vspace{10pt}
 Zhong-Ying Fan$^{1\dagger}$\\

\vspace{10pt}
$^{1\dagger}${ Department of Astrophysics, School of Physics and Material Science, \\
 Guangzhou University, Guangzhou 510006, China }\\


\vspace{40pt}

\underline{ABSTRACT}
\end{center}

We study upper bounds on the growth of operator entropy $S_K$ in operator growth. Using uncertainty relation, we first prove a dispersion bound on the growth rate $|\partial_t S_K|\leq 2b_1 \Delta S_K$, where $b_1$ is the first Lanczos coefficient and $\Delta S_K$ is the variance of $S_K$. However, for irreversible process, this bound generally turns out to be too loose at long times. We further find a tighter bound in the long time limit using a universal logarithmic relation between Krylov complexity and operator entropy. The new bound describes the long time behavior of operator entropy very well for physically interesting cases, such as chaotic systems and integrable models.

\vfill {\footnotesize  Email: $^\dagger$fanzhy@gzhu.edu.cn\,,}

\thispagestyle{empty}

\pagebreak

\tableofcontents
\addtocontents{toc}{\protect\setcounter{tocdepth}{2}}




\section{Introduction}

  The Krylov complexity ( or K-complexity) \cite{Parker:2018yvk} and the operator entropy (or K-entropy) \cite{Barbon:2019wsy} are introduced to describe the Heisenberg evolution of operators $\mc{O}(t)=e^{iHt}\mc{O}_0 e^{-iHt}$, where $\mc{O}_0$ is an initial operator and $H$ is the Hamiltonian. In particular, for complex systems, the initially simple operator $\mc{O}_0$ will irreversibly grow into a complex one, with the size blowing up exponentially. It is widely believed that a statistical description should emerge in this process and some universal features may be captured by K-complexity and K-entropy.

 The two notions have attracted a lot of attentions in literature \cite{Caputa:2021sib,Jian:2020qpp,Rabinovici:2020ryf,Dymarsky:2021bjq,Kim:2021okd,Patramanis:2021lkx,Adhikari:2022whf,Fan:2022xaa,Hornedal:2022pkc,Balasubramanian:2022tpr}. In particular, it was established in \cite{Fan:2022xaa} that for general irreversible process, the two quantities enjoy a universal logarithmic relation to leading order at long times: $ S_K=\eta\,\mm{ln}{K}+\cdots $, where $0<\eta\leq 1$ and $\eta=1$ corresponds to chaotic systems\footnote{By chaotic systems, we mean the Lanczos coefficients grows linearly asymptotically $b_n\rightarrow \alpha n+\gamma$, except for $1\mm{d}$ case, in which there is a logarithmic correction $b_n\rightarrow \alpha n/\mm{ln}{n}$. However, the authors in \cite{Dymarsky:2021bjq} show that for free field theories at finite temperatures, which do not probe chaos dynamically, the Lanczos coefficients grow linearly asymptotically as well. This raises a question whether asymptotic behavior of Lanczos coefficients characterizes dynamical properties of Hamiltonian systems appropriately. Nonetheless, this does not influence our discussions and results.}. On the other hand, it was proved in \cite{Hornedal:2022pkc} that the growth rate of K-complexity in operator growth is upper bounded as $|\partial_t K|\leq 2b_1 \Delta K$, where $b_1$ is the first Lanczos coefficient and $\Delta K$ is the variance of K-complexity. This is referred to as {\it dispersion bound} \cite{Hornedal:2022pkc}. This inspires us to search a similar bound for K-entropy and examine the influence of the above logarithmic relation to the growth of K-entropy.

The paper is organized as follows. In section 2, we briefly review the recursion method and the Lanczos algorithm. In section 3, we prove a upper bound on the growth of general A-complexity defined in the Kryolv space. We show that saturation of the bound demands that A-complexity operator is linearly related to K-complexity operator. In section 4, we study the upper bound on the growth of K-entropy. Using continuum limit, we establishes that the dispersion bound of K-entropy is rarely saturated. Instead, for irreversible process, the logarithmic relation between K-complexity and K-entropy implies a tighter bound in the long time limit. We test the new bound using a number of numerical examples. We conclude in section 5.

\section{The recursion method and the Lanczos algorithm}

For a lattice system with Hamiltonian $H$, the initial operator $\mc{O}_0$ evolves in time according to
\be \mc{O}(t)=e^{iHt}\mc{O}_0 e^{-iHt}=\sum_{n=0}\fft{(it)^n}{n!}\tilde{\mc{O}}_n\,, \ee
where $\tilde{\mc{O}}_n$ stands for the nested commutators: $\tilde{\mc{O}}_1=[H,\mc{O}_0]\,,\tilde{\mc{O}}_2=[H,\tilde{\mc{O}}_1]\,,\cdots\,,\tilde{\mc{O}}_n=[H,\tilde{\mc{O}}_{n-1}]\,.$
However, sometimes evaluation of these commutators is of great difficulty. It is helpful to take the operator as a wave function, which evolves under the {\it Liouvillian} $\mc{L}\equiv [H\,,\cdot]$. One has
\be\label{obasis} |\mc{O}(t))=e^{i\mc{L}t}|\mc{O}_0 )= \sum_{n=0}\fft{(it)^n}{n!}|\tilde{\mc{O}}_n )\,, \ee
where $|\tilde{\mc{O}}_n )=\mc{L}^n|\mc{O}_0 ).$ To proceed, one needs introduce an inner product in the operator Hilbert space satisfying
\be ( A|A )\geq 0\,,\quad ( A|B )=(B|A)^* \,,\quad (A|aB)=a (A|B) \,,\ee
where $a$ is a complex number. In addition, the Liouvillian should be hermitian under the inner product $(A|\mc{L}B)=(\mc{L}A|B)$. It is clear that the choice of inner product greatly influences the outcome of the recursion method, such as the behavior of Lanczos coefficient $b_n$ introduced below. However, for our purpose in this paper, we do not need to specify a particular choice of inner product ( when we discuss certain Hamiltonian systems, the inner product generally has already been specified in literature but we will not mention it explicitly. The readers should not be confused). The interested readers are referred to \cite{vsmuller} for more details about choice of inner product.

The physical information about operator growth is essentially encoded in the auto-correlation function
\be C(t):= ( \mc{O}_0|\mc{O}(t) )=( \mc{O}_0| e^{i\mc{L}t}  |\mc{O}_0 )\,. \ee

In general, the original operator basis $\{|\tilde{\mc{O}}_n )\}$ is not orthogonal. To study the operator dynamics in an orthonormal basis, one adopts the Gram-Schmidt scheme, starting with a normalized vector $|\mc{O}_0 )$. The first vector is given by $|\mc{O}_1):=b_1^{-1}\mc{L}|\mc{O}_0)$, where $b_1:=(\mc{O}_0\mc{L}|\mc{L}\mc{O}_0 )^{1/2}$. For the $n-$th vector, one has inductively
\bea\label{kbasis}
&&|A_n):=\mc{L}|\mc{O}_{n-1})-b_{n-1}|\mc{O}_{n-2})\,,\nn\\
&&|\mc{O}_n ):=b_n^{-1}|A_n )\,,\quad b_n:=( A_n|A_n )^{1/2}\,.
\eea
The output of this procedure is a set of orthomornal vectors $\{|\mc{O}_n )\}$, referred to as {\it Krylov basis} and a sequence of positive numbers $\{b_n\}$, referred to as {\it Lanczos coefficients}. These coefficients have units of energy and can be used to measure time in the Heisenberg evolution of operators.

In Krylov basis, evolution of the operator $\mc{O}(t)$ can be formally written as
\be\label{newbasis} |\mc{O}(t)):=\sum_{n=0}i^n\varphi_n(t)|\mc{O}_n)\,,  \ee
where $\varphi_n(t):=i^{-n}(\mc{O}_n|\mc{O}(t))$ is a discrete set of (real) wave functions and $p_n\equiv \varphi_n^2$ can be interpreted as probabilities. One has the normalization
\be \sum_{n=0}^{\infty}\varphi^2_n(t)=1 \,.\ee

The Heisenberg evolution of $\mc{O}(t)$ gives rise to a discrete set of equations
\be\label{varphi} \partial_t\varphi_n=b_n\varphi_{n-1}-b_{n+1}\varphi_{n+1} \,,\ee
subject to the boundary condition $\varphi_n(0)=\delta_{n0}$ and $b_0=0=\varphi_{-1}(t)$ by convention. This uniquely determines the wave functions $\varphi_n(t)$ for a given set of Lanczos coefficients. Since the auto-correlation function is simply $C(t)=\varphi_0(t)$, it is immediately seen that physical information encoded in $C(t)$ can be equivalently extracted from the Lanczos coefficients, though there is not a simple transformation between them.

The Krylov complexity (K-complexity) and the Krylov entropy (K-entropy) are defined as
\bea  K&=&( \mc{O}(t)|\mc{K}| \mc{O}(t))\,,\quad \mc{K}=\sum_{n=0}n |\mc{O}_n) ( \mc{O}_n| \,,\\
     S_K&=& ( \mc{O}(t)|\mc{S}| \mc{O}(t) )\,,\quad \mc{S}=\sum_{n=0}\mc{S}_n\, |\mc{O}_n) ( \mc{O}_n| \,,\eea
where $\mc{S}_n\equiv -\,\mm{ln}{\varphi_n^2}$. Generalisation of K-complexity is also interesting, for example the K-complexity of degree $i$ is defined as
\be  K_i=( \mc{O}(t)|\mc{K}_{(i)}| \mc{O}(t) )\,,\quad \mc{K}_{(i)}=\sum_{n=0}n^i |\mc{O}_n ) ( \mc{O}_n| \,. \ee
These quantities roughly play the same role as K-complexity in characterizing operator growth.
However, except for simplicity, the definition of K-complexity is special in the sense that it is deeply connected to the complexity algebra, which characterizes the symmetry of Heisenberg evolution \cite{Hornedal:2022pkc}. We will study this in the next section. Here we just remind the readers that if and only if the complexity algebra is closed, the operator wave functions can be exactly solved and the dispersion bound on the growth rate of K-complexity is saturated \cite{Hornedal:2022pkc}.

\section{Upper bound on the growth of A-complexity}

In \cite{Hornedal:2022pkc}, the authors first proved an upper bound on the growth rate of K-complexity, referred to as {\it dispersion bound}. It was shown that saturation of the bound is equivalent to closure of the complexity algebra, which is spanned by $\{ \mc{L}_+\,,\mc{L}_-\,,\mc{B}=[\mc{L}_-\,,\mc{L}_+] \}$, where
\bea &&\mc{L}_+=\sum_{n=0}b_{n+1}\,|\mc{O}_{n+1}) (\mc{O}_n|\,,\nn\\
&&\mc{L}_-=\sum_{n=0}b_{n+1}\,|\mc{O}_{n} ) (\mc{O}_{n+1}|\,.\eea
Notice the Liouvillian $\mc{L}=\mc{L}_++\mc{L}_-$. In fact, only when complexity algebra is closed, the operator wave functions $\varphi_n$ can be solved exactly \cite{Caputa:2021sib} and hence these cases provide interesting examples to study operator growth. In this section, we would like to extend the result \cite{Hornedal:2022pkc} to general A-complexity defined under the Krylov basis
\be A(t)\equiv (\mc{O}(t)|\mc{A}| \mc{O}(t))\,,\quad \mc{A}=\sum_{n=0}\mc{A}_n |\mc{O}_n ) ( \mc{O}_n|\,,\ee
where $\mc{A}$ is a self-adjoint superoperator. We demand that $\mc{A}_n$ is either a polynomial of $n$, independent of time or $\mc{A}_n=\mc{S}_n$ when it depends on time. Clearly K-complexity and K-entropy are included in our notions of A-complexity. In the following, we will generalise the dispersion bound of K-complexity to this entire class of complexity.

First, by straightforward calculations, we deduce
\bea\label{15}
&&( \mc{O}(t)|\mc{A}\mc{L}|\mc{O}(t))=\sum_{n=0}-i\mc{A}_n \varphi_n \dot{\varphi}_n \,,\nn\\
&&(\mc{O}(t)|\mc{L}\mc{A}|\mc{O}(t))=\sum_{n=0}i\mc{A}_n \varphi_n \dot{\varphi}_n \,,
\eea
where a dot denotes the derivative with respect to the canonical time coordinate. This gives
\be( \mc{O}(t)|[\mc{A}\,,\mc{L}]|\mc{O}(t) ) = -i\sum_{n=0}\mc{A}_n  \dot{p}_n  \,.\ee
On the other hand, according to definition, the growth rate of A-complexity is given by
\be \partial_t A(t)=\sum_{n=0}\Big[\mc{A}_n +p_n\partial_{p_n}\mc{A}_n \Big] \dot{p}_n  \,.\ee
If $\mc{A}_n$ is independent of time, the second term in the square bracket vanishes whereas if $\mc{A}_n=\mc{S}_n$, this term still does not contribute because of $\sum_{n=0}\dot{p}_n=0$. In both cases, we arrive at
\be -i\partial_t A=( \mc{O}(t)|[\mc{A}\,,\mc{L}]|\mc{O}(t) )\,.\ee
In fact, the result can be derived more directly by taking the operator space as an ordinary Hilbert space, which evolves under the ``Hamiltonian" $\mc{L}$. Using the density operator $\rho\equiv |\mc{O}(t)) (\mc{O}(t)| $ and $A=\mm{Tr}(\rho\mc{A})$, one finds
\be -i\partial_t A=-i\mm{Tr}(\dot\rho \mc{A})=\mm{Tr}([\mc{L}\,,\rho]\mc{A})=(\mc{O}(t)|[\mc{A}\,,\mc{L}]|\mc{O}(t)) \,,\ee
where in the second equality, we have adopted the Liouville equation $\dot\rho=i[\mc{L}\,,\rho]$.

 With this result in hand and adopting the Robertson uncertainty relation $2\Delta A\Delta\mc{L}\geq |\langle [\mc{A}\,,\mc{L}] \rangle|$, we obtain
\be |\partial_t A|\leq 2 \Delta A\Delta\mc{L}=2b_1\Delta \mc{A}\,,\label{boundgene}\ee
where $\Delta A=\sqrt{\langle\mc{A}^2\rangle-\langle \mc{A} \rangle^2}$ stands for the dispersion of $\mc{A}$ with respect to some state $|A)$ and $\Delta\mc{L}=b_1$. Notice that if the Krylov space is infinite dimensional, it is necessary that the state $|A)$ is contained in the intersection of the domains of $\mc{A}\mc{L}$ and $\mc{L}\mc{A}$, otherwise the bound may not hold \cite{davidson}. To avoid confusion with the uncertainty relation with observables, the bound is referred to as {\it dispersion bound} in \cite{Hornedal:2022pkc} and we shall follow this convention.

Several comments are in order:

$\bullet$ The bound is valid to any kind of A-complexity, including K-complexity and K-entropy. These two cases will be studied carefully later.

$\bullet$ The initial time $t=0$ is an extremal point, at which the bound is always saturated for any A-complexity.

$\bullet$ A geometrical interpretation was given for saturation of the bound for K-complexity \cite{Hornedal:2022pkc}: The projective Krylov space can be identified with all rank one orthogonal projections on the Krylov space. The bound is saturated if and only if the corresponding curve in the Krylov space moves along the gradient of Krylov complexity: meaning the dynamics is directed to the direction that maximizes the local growth of K-complexity. This interpretation can be generalised to any kind of A-complexity straightforwardly.

$\bullet$ There are several cases, in which the complexity algebra is closed. It was proved \cite{Hornedal:2022pkc} that only in these cases the dispersion bound for K-complexity is saturated. Later, we will show that for general A-complexity, saturation of the dispersion bound is possible if and only if $\mc{A}$ is linearly related to $\mc{K}$.

$\bullet$ One can not obtain a tighter bound by using generalised uncertainty relation because of $\langle \{ \mc{A}\,,\mc{L} \} \rangle=0$, see (\ref{15}). However, this does not exclude the possibility that for certain cases (such as irreversible process), the growth rate of A-complexity could be tighter bounded at some time regimes. We will study this in detail for K-entropy in the next section.

\subsection{Conditions for saturation of the dispersion bound}

Apparently it is not clear whether the dispersion bound of A-complexity can be saturated during unitary evolution of operators. However, since the uncertainty relation is obtained by applying the Cauchy-Schwarz inequality to the two vectors $(\mc{A}-A)|\mc{O}(t))$ and $(\mc{L}-\langle \mc{L}\rangle)|\mc{O}(t))$, saturation of the bound is equivalent to the two vectors are linearly dependent. In other words, the vector $e^{-i\mc{L}t}(\mc{A}-A)e^{i\mc{L}t}|\mc{O}_0)$ is linearly related to $|\mc{O}_1)$ because of $\langle \mc{L}\rangle=0$. To study this carefully, we adopt the Baker-Campbell-Hausdorff formula and find
\be  e^{-i\mc{L}t}(\mc{A}-A)e^{i\mc{L}t}|\mc{O}_0 )=\Big[ -A(t)+\sum_{n=0}\fft{(-it)^n}{n!}L^n    \Big]|\mc{O}_0 ) \,,\label{Acomp}\ee
where $L^n$ stands for the nested commutators: $L^0=\mc{A}\,,L^1=[\mc{L}\,,\mc{A}]\,,L^2=\big[ [\mc{L}\,,\mc{A}]\,,\mc{A}\big]\,,\cdots\,.$

To evaluate these commutators, we introduce matrix basises $e_{m\,,n}=|\mc{O}_m ) ( \mc{O}_n|$ and represent a matrix as $M=\sum M_{m\,,n}e_{m\,,n}$. Here $e_{n\,,n}$ corresponds to the diagonal elements. For later purpose, we refer the elements associated to the basis $e_{n+k,n}$ (or $e_{n,n+k}$) as $k$-diagonals (or $-k$-diagonals) of operators. One has
\be e_{m\,,n}|\mc{O}_k)=\delta_{nk}|\mc{O}_m) \,,\quad e_{m\,,p}\,e_{q\,,n}=\delta_{pq}e_{m\,,n}\,.\ee
We are ready to compute the nested commutators $L^n$:

$\bullet$ First, $L^0=\mc{A}$ and $L^0|\mc{O}_0 )=\mc{A}_0 |\mc{O}_0 )$. Similar terms giving $|\mc{O}_0)$ are widely contained in the nested commutators $L^n$. In fact, if the upper bound is saturated, these terms will be cancelled by $A(t)$ term in (\ref{Acomp}). It provides an alternate way to extract A-complexity in this case, see \cite{Hornedal:2022pkc} for K-complexity case.

$\bullet$ Second, straightforward calculation gives $L^1=\sum_{n=0}b_{n+1}(\mc{A}_n-\mc{A}_{n+1})(e_{n+1\,,n}-e_{n\,,n+1})$
and $L^1|\mc{O}_0)=b_1(\mc{A}_0-\mc{A}_1)|\mc{O}_1)$. This gives the desired term proportional to $|\mc{O}_1)$. However, evaluation of $L^2$ yields
\be L^2=\sum_{n=0}g(n)b_{n+1}b_{n+2}(e_{n+2\,,n}+e_{n\,,n+2})+\cdots \,,\ee
where $g(n)=\mc{A}_n-\mc{A}_{n+1}-(\mc{A}_{n+1}-\mc{A}_{n+2})$ and we have omitted the diagonal terms, which are irrelevant to our discussions. One finds $L^2|\mc{O}_0 )=g(0)b_1b_2|\mc{O}_2 )+\cdots$. Obviously this term cannot be cancelled by the A-complexity term and hence must be set to zero for saturation of the upper bound.

$\bullet$  As a matter of fact, the highest subdiagonal in $L^n$ is $n$-diagonal (or $-n$-diagonal) so that $L^n|\mc{O}_0)=(L^n)_{n,0}|\mc{O}_n)+\cdots$. Saturation of the upper bound demands that all these terms vanish. It follows that the $(n+2)$-diagonals of $L^{n+2}$ can be deduced from the $(n+1)$-diagonals of $L^{n+1}$:
\be L(n\,,k)=b_{n+k+2}\,L(n-1\,,k)-b_{k+1}\,L(n-1\,,k+1) \,,\ee
where $n\geq 1$ and $L(n\,,k)\equiv \big( L^{n+2}\big)_{n+k+2\,,k}$. Mathematically for above recurrence relations, together with the initial condition $L(0\,,k)=g(k)b_{k+1}b_{k+2}$, one has \cite{Hornedal:2022pkc}
\be L(n\,,k)=\prod_{j=k+1}^{n+k+2}b_j \sum_{q=0}^n (-)^q \binom{n}{k} g(k+q) \,.\ee
This implies that $g(k)=0$ for $k\geq 1$ to guarantee all $n$-diagonals of $L^n$ vanishes. Hence, $\mc{A}_n$ must be a linear function of $n$: $\mc{A}_n=\alpha n+\beta$, where $\alpha\,,\beta$ are constants independent of $n$ (but may depend on time). In other words, saturation of the upper bound for A-complexity is possible when A-complexity operator $\mc{A}$ is linearly related to the K-complexity operator $\mc{A}=\alpha \mc{K}+\beta$. Since the bound for K-complexity is saturated if and only if the complexity algebra is closed \cite{Hornedal:2022pkc}, it is a little surprising that the algebra influences the unitary evolution of operators more widely than expected: it constrains the growth rate of any kind of A-complexity.

\section{Bound on the growth of K-entropy}

\subsection{Saturation of the dispersion bound}
 We move to study K-entropy. According to previous discussions, saturation of the dispersion bound on K-entropy is only possible when the complexity algebra is closed and hence the bound on K-complexity is saturated. These are only a few cases \cite{Caputa:2021sib,Hornedal:2022pkc}: the complexity algebra is $\mm{SL}(2\,,\mathbb{R})$ algebra (the SYK model), $\mm{SU}(2)$ algebra and Heisenberg-Weyl algebra. Exploring these cases, we only find two examples which saturate the K-entropy bound.

The first is the evolution of operator $\mc{O}_0=\sigma_x$ under a single qubit (two-level) Hamiltonian $H=\omega\sigma_z$, where $\sigma_i$ stands for the $i$-th Pauli matrix. This case belongs to $j=1/2$ case of the $\mm{SU}(2)$ algebra. One has $b_1=\omega\,,b_{n>1}=0\,,$ and $\varphi_0=\cos(\omega t)\,,\varphi_1=\sin(\omega t)$. However, since the Krylov dimension is only $D=2$, it turns out that the dispersion bound for any A-complexity is trivially saturated\footnote{In addition to the K-entropy, consider $A=\sum_n P(n)\varphi_n^2(t)$, where $P(n)$ is a polynomial of $n$, independent of time. It is easy to see $A=P(1)\varphi_1^2(t)\,,\Delta A=|P(1)\varphi_0(t)\varphi_1(t)|$ so that $|\partial_tA|=2b_1\Delta A$.}. For example, evaluation of K-entropy yields
\bea
&&S_K=-\cos^2(\omega t)\,\mm{ln}\,{\cos^2(\omega t) }-\sin^2(\omega t)\,\mm{ln}\,{\sin^2(\omega t) }\,,\nn\\
&&\Delta S_K=2\sin(\omega t)\cos(\omega t)\,\mm{ln}\,{\cot(\omega t)}\,.
\eea
It is easy to see $|\partial_t S_K|=2b_1 \Delta S_K$.

The second example belongs to a special case of SYK model. The Lanczos coefficient is proportional to $n$: $b_n=\omega n$, with the wave functions given by $\varphi_n(t)=\tanh^n{(\omega t)}/\cosh{(\omega t)}$.
It is easy to see that for this case the K-entropy operator is linearly related to K-complexity operator. Explicitly, the K-entropy and the variance are given by
\bea\label{syk1}
&& S_K=\cosh^2(\omega t)\,\mm{ln}\,{\cosh^2(\omega t) }-\sinh^2(\omega t)\,\mm{ln}\,{\sinh^2(\omega t) } \,,\nn\\
&& \Delta S_K=2\sinh(\omega t)\cosh(\omega t)\,\mm{ln}\,{\coth(\omega t)}\,.
\eea
Again $|\partial_t S_K|=2b_1 \Delta S_K$. It is interesting to observe that by demanding reality and positivity, the above results for K-entropy (and K-complexity) can be obtained by taking $\omega\rightarrow i\omega$ from the single-qubit system. This is unexpected. However, the trick is not valid to general A-complexity.

\subsection{Continuum limit analysis}\label{continuum}

From above discussions, one may gain an intuition that saturation of the dispersion bound for K-entropy in the unitary evolution of operators is much harder than that for K-complexity. In fact, for the latter, the bound is always asymptotically saturated at long times for chaotic systems\footnote{The condition for saturation of the dispersion bound of K-complexity is $b_{n+1}^2-b_n^2$ is linear in $n$ whereas for chaotic systems the Lanczos coefficient always grows asymptotically linearly. Hence, the bound is saturated asymptotically at long times for general chaotic systems.} but this never happens for K-entropy (or any other A-complexity). To search a deeper understanding for this, we adopt the continuum limit to study the dispersion bound for the two quantities.

It was known that for semi-infinite chains, the continuum limit is good at capture the leading long time behaviors of K-complexity and K-entropy using coarse grained wave functions \cite{Barbon:2019wsy}.
Introducing a lattice cutoff $\epsilon$ and defining a coordinate $x=\epsilon n$ and the velocity $v(x)=2\epsilon b_n$. The interpolating wave function is defined as $\varphi(x\,,t)=\varphi_n(t)$. The wave equation (\ref{varphi}) becomes
\be \partial_t\varphi(x\,,t)=\fft{1}{2\epsilon}\Big[ v(x)\varphi(x-\epsilon)-v(x+\epsilon)\varphi(x+\epsilon) \Big] \,. \ee
Expansion in powers of $\epsilon$, one finds to leading order
\be\label{expansion} \partial_t\varphi=-v(x)\partial_x\varphi-\fft12 \partial_x v(x)\varphi+O(\epsilon) \,.\ee
This is a chiral wave equation with a position-dependent velocity $v(x)$ and mass $\fft12 \partial_x v(x)$. The equation is much simplified in a new frame $y$ defined as $v(x)\partial_x=\partial_y$ and a rescaled wave function
\be \psi(y\,,t)=\sqrt{v(y)}\,\varphi(y\,,t) \,.\ee
One finds
\be\label{chiralwave} (\partial_t+\partial_y)\psi(y\,,t)=0+O(\epsilon) \,.\ee
The general solution is given by
\be \psi(y\,,t)=\psi_i(y-t) \,,\ee
where $\psi_i(y)=\psi(y\,,0)$ stands for the initial amplitude. This tells us that the leading order wave function simply moves ballistically in the evolution. This approximation derives the growth of K-complexity correctly but for K-entropy, some higher order corrections should be included for certain cases, for example the Lanczos coefficients have a bounded support \cite{Barbon:2019wsy}. However, for our purpose, the leading order result is already enough (we will turn to numerical approach when the leading order analysis breaks down ).

Normalization condition reads
\be 1=\sum_{n}|\varphi_n(t)|^2=\fft{1}{\epsilon}\int \mm{d}x\, \varphi^2(x\,,t)=\fft{1}{\epsilon}\int \mm{d}y\, \psi^2(y\,,t)  \,.\ee
Evaluation of the K-complexity and the K-entropy yields
\bea\label{cksk}
K(t)&=&\sum_{n}n |\varphi_n(t)|^2=\fft{1}{\epsilon}\int\mm{d}y\, \fft{x(y+t)}{\epsilon}\,\, \psi_i^2(y) \,,\\
S_K(t)&=&-\sum_{n}|\varphi_n(t)|^2\,\mm{ln}{|\varphi_n(t)|^2} \nn\\
      &=&\fft{1}{\epsilon}\int\mm{d}y\, \psi_i^2(y)\Big[\,\mm{ln}{v(y+t)}-\,\mm{ln}{\psi_i^2(y)}\Big] \,.\eea
Similarly, one has
\bea
K_2(t)&=&\sum_{n}n^2 |\varphi_n(t)|^2=\fft{1}{\epsilon}\int\mm{d}y\, \fft{x^2(y+t)}{\epsilon^2}\,\, \psi_i^2(y) \,,\\
S_{K_2}(t)&=&\sum_{n}|\varphi_n(t)|^2\,\mm{ln}^2{|\varphi_n(t)|^2} \nn\\
          &=&\fft{1}{\epsilon}\int\mm{d}y\, \psi_i^2(y)\Big[\,\mm{ln}{v(y+t)}-\mm{ln}{\psi_i^2(y)}\Big]^2 \,.
\eea
Using these results, once the transformation between the two frames is known, we are able to extract the leading time dependence of the quantities immediately.

\begin{figure}
  \centering
  \includegraphics[width=300pt]{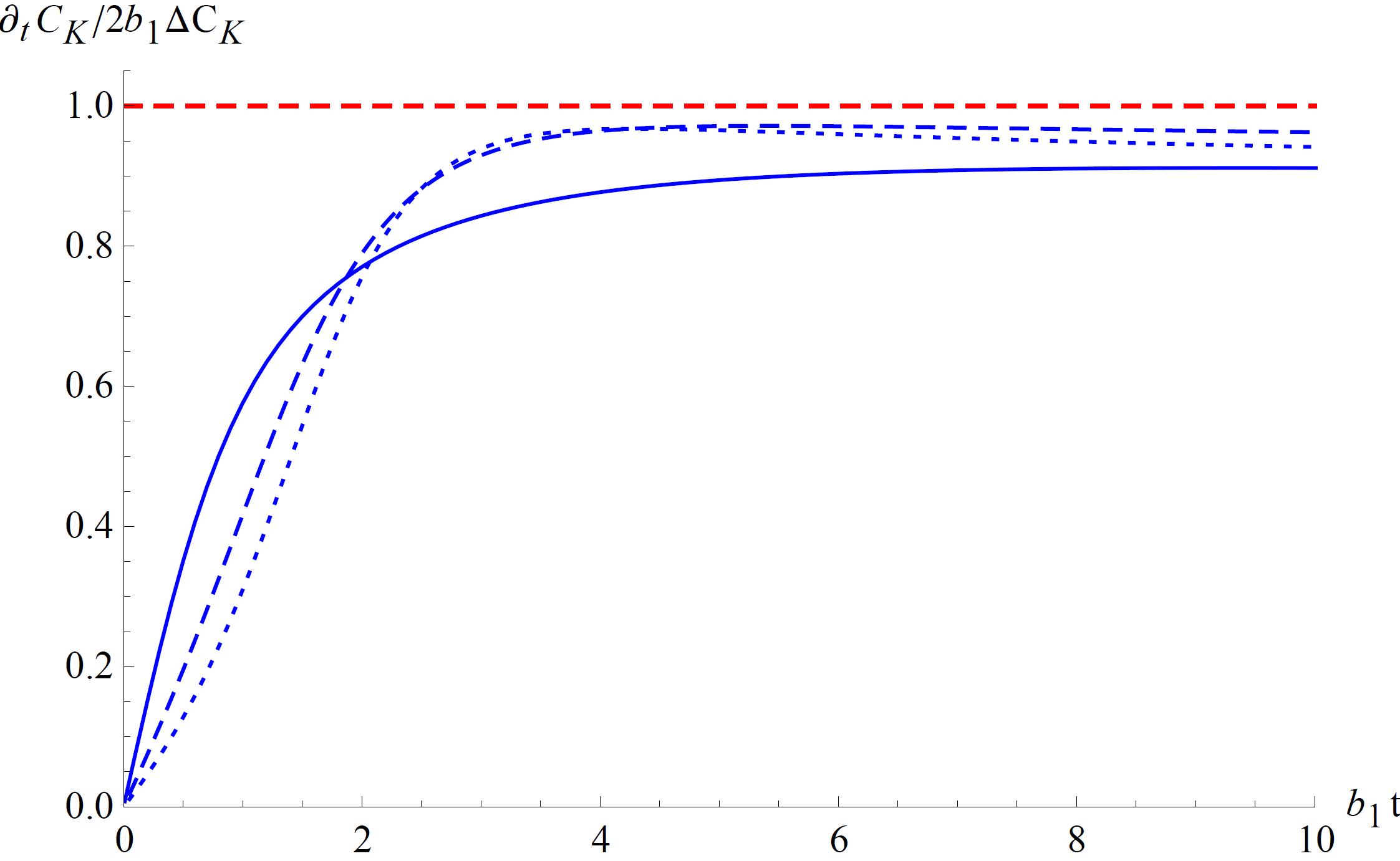}
  \caption{The ratio of the growth rate of K-complexity to the dispersion bound in operator growth for integrable models $b_n=\alpha n^\delta$, where $\delta=3/8$ (solid), $2/3$ (dashed) and $3/4$ (dotted). The horizontal line stands for the cases in which the bound is saturated during the evolution. It is immediately seen that even if for these models, the bound is not saturated asymptotically, complexity still grows fast, with the rate of change in the same order of the dispersion bound.}
  \label{kbound}
\end{figure}

Here we are particularly interested in evaluating the variance of K-complexity and K-entropy. Using Taylor expansion, we deduce to leading order at long times
\be \Delta K\simeq \fft{x'(t)}{\epsilon}\sqrt{Y_2-Y_1^2}\sim \partial_t K/b_1 \,,\ee
where $Y_n\equiv \fft{1}{\epsilon}\int dy\,y^n\,\psi^2_i(y)$ and we have ignored higher order terms (these terms are in the same order of $x'(t)/\epsilon$ for chaotic systems but this does not influence our discussions). This implies that while saturation of the dispersion bound happens only for a few cases, the bound still captures the long time behavior of K-complexity well. In other words, even if the dispersion bound is not saturated asymptotically in operator growth, the K-complexity still grows fast, with a rate of change close to the dispersion bound. This is also supported by our numerical results. For example, in Fig.\ref{kbound}, we show this explicitly for integrable models $b_n=\alpha n^\delta$.

The variance of K-entropy can be analyzed in the same way. We find to leading order
\be \Delta S_K=\fft{1}{\epsilon}\int\mm{d}y\, \psi_i^2(y)\,\mm{ln}^2{\psi_i^2(y)}-\fft{1}{\epsilon^2}\Big[\int\mm{d}y\, \psi_i^2(y)\,\mm{ln}{\psi_i^2(y)}\Big]^2 \,,\label{deltask}\ee
which is of order unity. It implies that compared to K-complexity, the dispersion bound of K-entropy approaches to a constant at long times for general irreversible process. This is ensured by our numerical results as well (the interested readers are referred to the next subsection for details). In particular, for certain cases in which the leading order analysis breaks down, we find the result is still valid. Hence, we may take (\ref{deltask}) as a definite result for irreversible process.
However, the growth rate of K-entropy generally decreases in a power law at long times $\partial_t S_K\sim 1/t^\gamma$, where $0<\gamma\leq 1$ \cite{Fan:2022xaa} except in the scrambling regime of chaotic systems (where $\partial_t S_K$ is a constant). This illustrates technically why the dispersion bound of K-entropy is too loose at long times\footnote{However, for chaotic systems, $\partial_t S_K$ is a constant, in the same order of the entropy variance $\partial_t S_K\sim 2b_1\Delta S_K$. Hence, the logic does not explain why in this case, the dispersion bound of K-entropy is not asymptotically saturated as K-complexity.}. This is also in accordance with physical expectations. At sufficiently long times, the distribution of wave functions at an instant $t$ nearly arrives at the maximal entropy (allowed by the operator dynamics) so that in the next time step $t\rightarrow t+\delta t$, the increasing of entropy is highly suppressed ($\delta S_K\sim \delta K/K$ according to (\ref{log})), while the dimension of the operator space, effectively characterized by K-complexity increases monotonically without any such constraint (this is said in the regime $0<K<S$, for systems with $S$ extensive degrees of freedoms).

The above analysis tells us that unlike K-complexity, the dispersion bound of K-entropy is too loose to characterize the long time behavior of K-entropy for irreversible process. This motivates us to search a tighter bound for the entropy growth. In the next subsection, we will show that such a bound indeed exists in the long time limit.

\subsection{A new bound on K-entropy in the long time limit}
It was established \cite{Fan:2022xaa} that for irreversible process, there exists a universal logarithmic relation between K-complexity and K-entropy to leading order at long times:
\be S_K=\eta \,\mm{ln}{K}+\cdots \,,\label{log}\ee
where the coefficient $\eta$ is bounded as $0<\eta\leq 1$. The upper bound $\eta=1$ is saturated by chaotic systems. In particular, it implies that long time dynamics of fast scramblers is particularly simple. Given the mean value of the distribution $D$, the wave functions to leading order are described by a uniform distribution so that the entropy is maximal. One has $S_K\simeq \log{D}$ and $K\simeq D/2$. Up to a factor of $2$, the K-complexity exactly measures the effective dimension of the Krylov space at an instant. It turns out that existence of the logarithemic relation bounds the growth of K-entropy further in the long time limit.

Consider K-complexity at first. On one hand, one has the dispersion bound $|\partial_t K|\leq 2b_1\Delta K$ whereas the logarithmic relation implies at long times
\be |\partial_t K|=K\,|\partial_t S_K|/\eta\leq 2b_1\,K\Delta S_K/\eta \,,\label{boundck}\ee
where in the first equality we have ignored the subleading order corrections. Hence, strictly speaking the result is valid only in the long time limit. All relevant results below should be understood in the same way. Now we have two bounds for complexity growth.
However, we have learnt from the continuum limit analysis that complexity growth rate can be close to the dispersion bound in the long time limit for irreversible process. Hence, it is hard to imagine that (\ref{boundck}) provides a tighter bound. Then consistency of the two bounds leads us to propose
\be t\rightarrow \infty\,,\quad |\partial_t K|\leq 2b_1\Delta K \leq 2b_1\,K\Delta S_K/\eta \,,\label{boundck2}\ee
or written more explicitly
\be t\rightarrow \infty\,,\quad \eta \Delta K/K\leq \Delta S_K \label{connection}\,.\ee
The relation can also be understood as the constant $\eta$ is bounded in the long time dynamics as
\be \eta\leq \lim_{t\rightarrow \infty}\Delta S_K K/\Delta K \label{eta}\,.\ee
Return to the growth of K-entropy. If the above result is correct, it implies a tighter bound on the K-entropy growth in the long time limit
\be t\rightarrow \infty\,,\quad |\partial_t S_K|\leq 2b_1\,\eta \Delta K/K\leq 2b_1\Delta S_K \,.\label{boundsk}\ee
This is a main result of this paper. We will show that though this new bound is valid only in the long time limit, it well describes the growth of K-entropy in the long time tails of irreversible process.

Before moving to explicit examples, let us show why (\ref{connection}) should be correct. First, consider fast scramblers. The K-complexity saturates the dispersion bound in the long time limit such that
\be \Delta K/K=\partial_t K/(2b_1 K)=\partial_t S_K/2b_1\leq \Delta S_K \label{connection2}\,.\ee
This proves (\ref{connection}). For general cases, $\Delta K\,,\Delta S_K$ at long times can be estimated using the continuum limit analysis. One has
$\Delta K/K\sim \partial_t K/b_1 K\sim \partial_t S_K/b_1$, which generally decreases (in a power law to leading order) at long times. However the variance of K-entropy approaches to a constant $\Delta S_K\sim O(1)$. Therefore, (\ref{connection}) continues hold beyond a critical time $t_c$ at which $\eta \Delta K(t_c)/K(t_c)=\Delta S_K(t_c)$. In fact, we obtain a stronger result
\be \eta\leq 1\leq  \lim_{t\rightarrow \infty}\Delta S_K K/\Delta K \label{eta2}\,.\ee
The relation passes a variety of numerical tests.


\subsubsection{Numerical results}

We would like to test the new bound of K-entropy proposed in (\ref{boundsk}) using a variety of numerical examples. Notice that (\ref{boundsk}) is equivalent to (\ref{connection}) or (\ref{eta}) so we shall focus on the former. Besides, it was shown in \cite{Fan:2022xaa} that the logarithmic relation (\ref{log}) begins to emerge around the time scale $t_*$, where $K(t_*)\sim O(1)$. Here $t_*$ coincides with the scrambling time for fast scramblers. It is interesting to see whether the result (\ref{boundsk}) can be extended to finite times $t\geq t_c$ as well.

\begin{figure}
  \includegraphics[width=230pt]{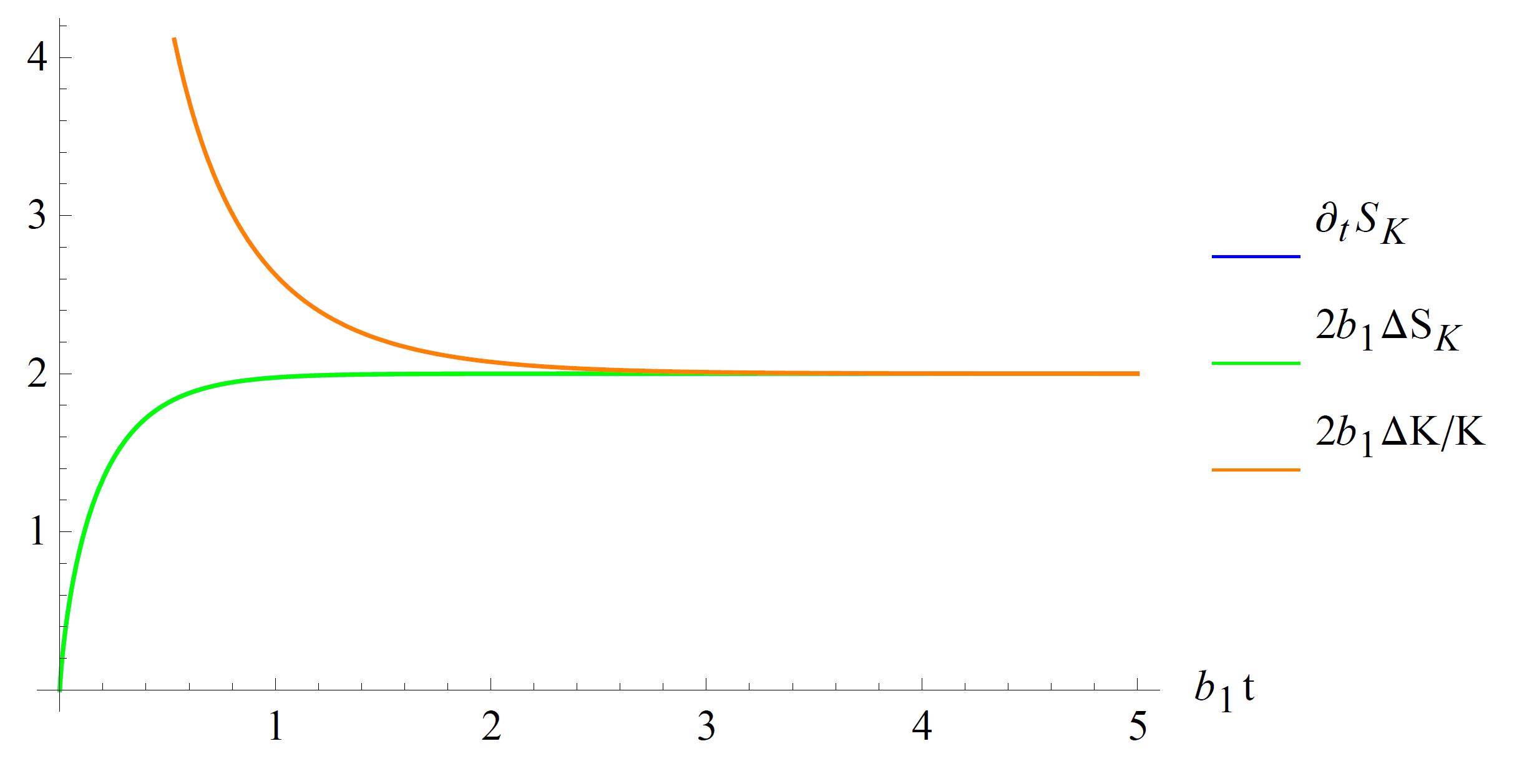}
  \includegraphics[width=230pt]{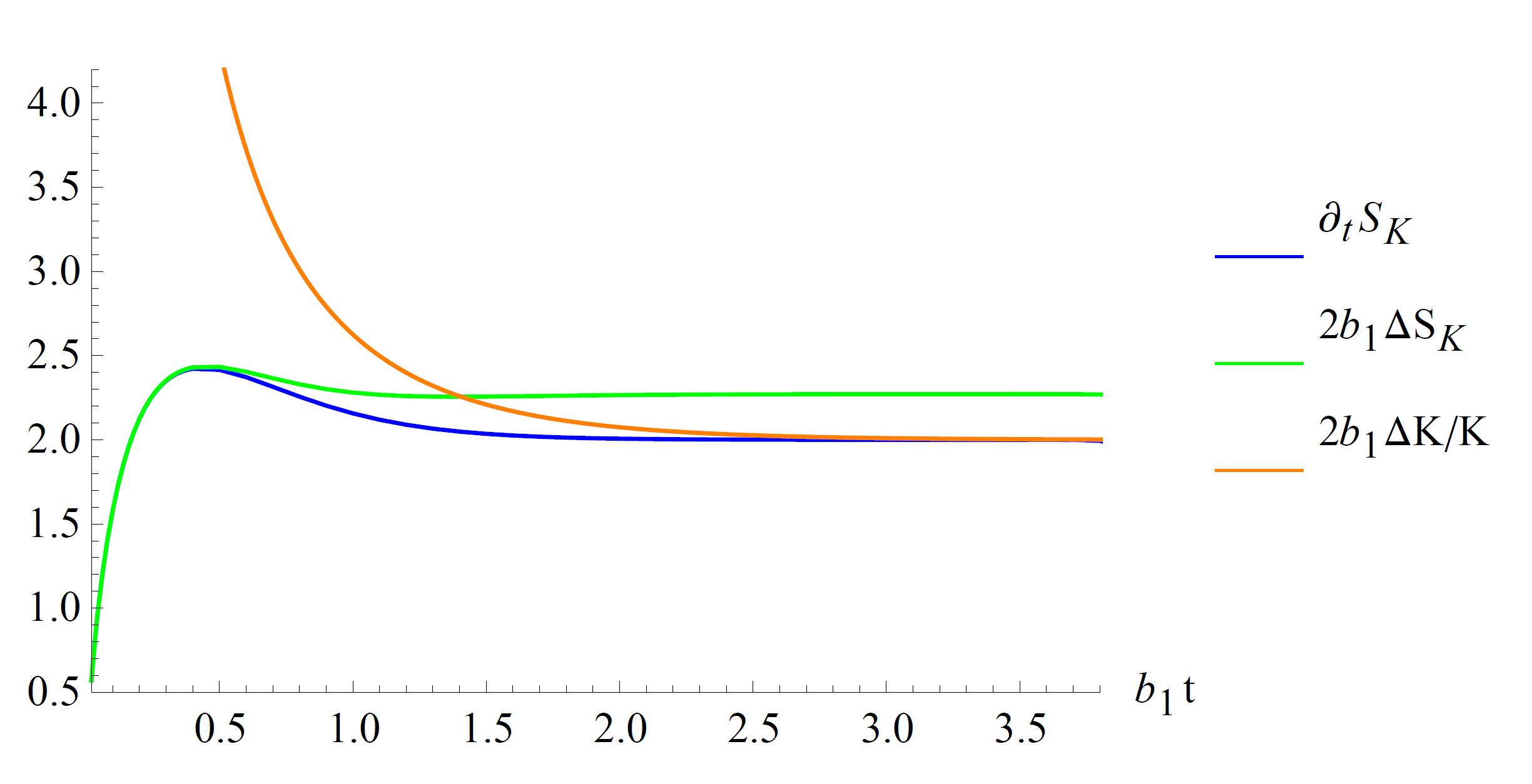}
  \caption{Entropy growth for SYK model. Left panel: $\xi=1$.  The bound $\partial_tS_K=2b_1 \Delta K/K$ is approached from below in the long time limit and $\partial_t S_K=2b_1\Delta K/K=2b_1 \Delta S_K$. However, during the evolution, the dispersion bound is always saturated and $\partial_t S_K=2b_1 \Delta S_K\leq 2b_1\Delta K/K$ at finite (but long) times. Right panel: $\xi=2$. The bound $\partial_tS_K=2b_1 \Delta K/K$ is again approached from below but the relation (\ref{boundsk}) can be extended to finite times $t\geq t_c$. }
  \label{syk}
\end{figure}

Consider SYK-like model at first. The Lanczos coefficient is given by $b_n=\omega \sqrt{n(n-1+\xi)}$. The wave functions can be solved exactly as \cite{Parker:2018yvk}
\be \varphi_n(t)=\sqrt{\fft{(\xi)_n}{n!}}\,\fft{\tanh^n(\omega t)}{\cosh^{\xi}{(\omega t)}}  \,,\ee
where $(\xi)_n=\xi(\xi+1)\cdots(\xi+n-1)$ is the Pochhammer symbol. For $\xi=1$, the K-entropy and its variance have already been given by (\ref{syk1}). It turns out that in this case, the bound (\ref{boundsk}) is approached from below in the long time limit: $\partial_t S_K= 2b_1\Delta K/K=2b_1 \Delta S_K$, as shown in the left panel of Fig.\ref{syk}. Here it is worth emphasizing that this is the only case we find in which the equality in (\ref{connection}) is taken. However, generalisation of the relation (\ref{boundsk}) to finite times turns out to be incorrect. Instead, since the dispersion bound is saturated during the evolution and $t_c\rightarrow \infty$, one finds $\partial_t S_K=2b_1 \Delta S_K\leq 2b_1\Delta K/K$.

The situation is much different for other $\xi$'s value. For example, the $\xi=2$ case is shown in the right panel of Fig.\ref{syk}. The tighter bound proposed in (\ref{boundsk}) is again approached from below but now $\partial_t S_K=2b_1\Delta K/K<2b_1\Delta S_K$ in the long time limit. In particular, the relation (\ref{boundsk}) can be extended to finite times $t\geq t_c$. As a matter of fact, we study a number of chaotic models numerically and find these features are generic.

\begin{figure}
  \includegraphics[width=230pt]{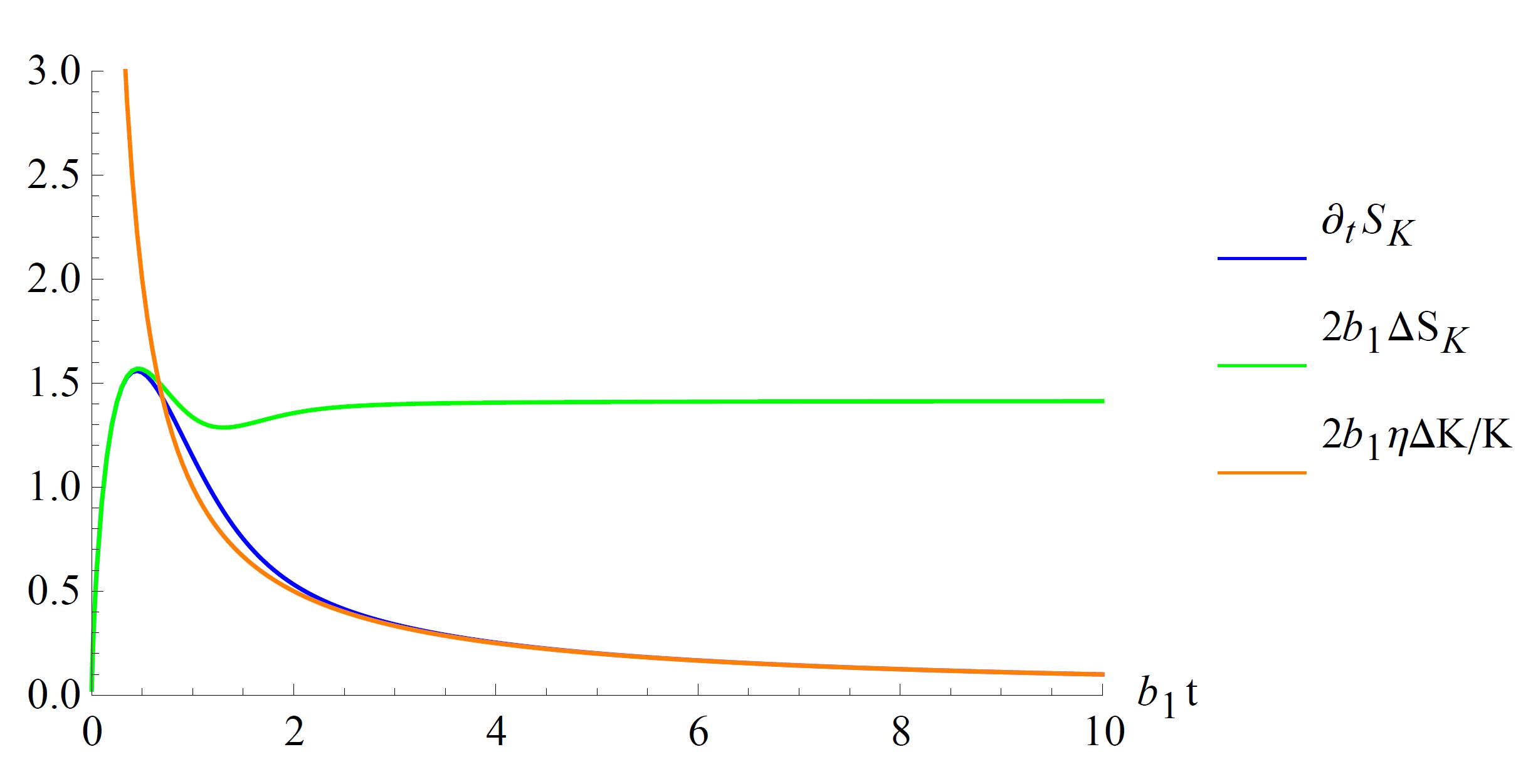}
  \includegraphics[width=230pt]{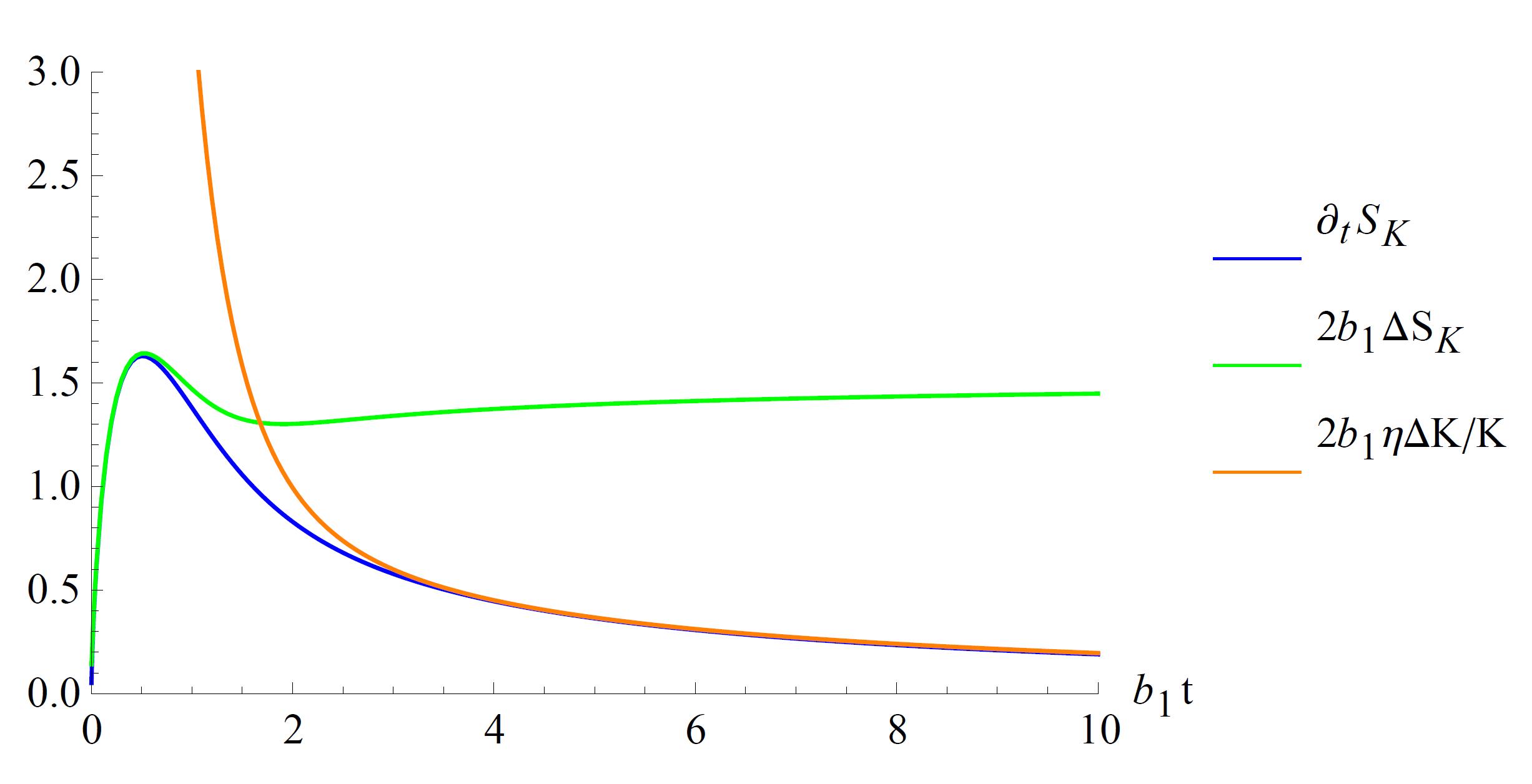}
  \caption{Entropy growth for integrable model $b_n=\alpha n^\delta$. Left panel: the Heisenberg-Weyl case $\delta=1/2$.  The bound $\partial_tS_K=2b_1 \Delta K/K$ is approached from above and $2b_1\eta \Delta K/K\leq \partial_tS_K<2b_1\Delta S_K$ beyond the critical time scale $t_c$. Right panel: $\delta=2/3$. The bound $\partial_tS_K=2b_1 \Delta K/K$ is approached from below and the relation (\ref{boundsk}) can be extended to finite times $t\geq t_c$. }
  \label{int}
\end{figure}

We move to study integrable models which have $b_n=\alpha n^\delta$, where $0<\delta<1$. In this case, the coefficient $\eta$ in the logarithmic relation (\ref{log}) is $\eta=\delta$ when $\delta \geq 1/2$ \cite{Fan:2022xaa}. In Fig.\ref{int}, we show the entropy growth during the evolution for two examples\footnote{For Heisenberg-Weyl case $\delta=1/2$, the wave functions can be solved exactly as $\varphi_n(t)=e^{-\alpha^2t^2/2}\fft{\alpha^n t^n}{\sqrt{n!}}$. }: $\delta=1/2$ (left) and $\delta=2/3$ (right). It follows that for integrable models, the growth rate of K-entropy generally decreases as $\partial_t S_K\sim 1/t$ whereas $\Delta S_K$ still approaches to a constant in the long time limit, consistent with our continuum limit analysis. Furthermore, the entropy growth rate is very close to the tighter bound (\ref{boundsk}) at long times. However, extension of relation (\ref{boundsk}) to finite times is not always correct. For instance, for the Heisenberg-Weyl case, the bound $\partial_tS_K=2b_1\eta \Delta K/K$ is approached from above so that the entropy growth rate is also bounded from below $2b_1\eta \Delta K/K\leq \partial_tS_K\leq 2b_1\Delta S_K$ at $t\geq t_c$ regime.

\begin{figure}
  \centering
  \includegraphics[width=300pt]{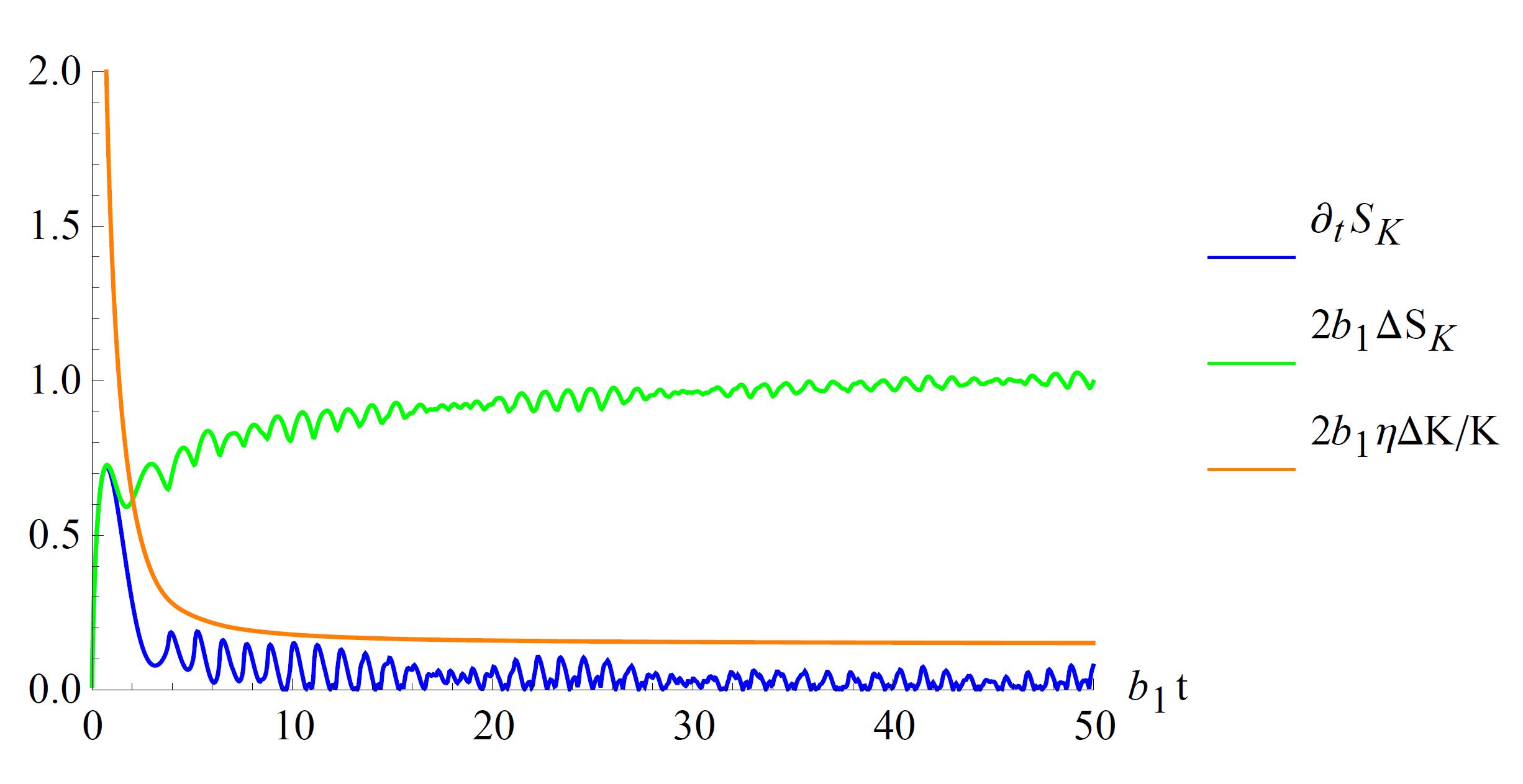}
  \caption{The K-entropy growth for the bounded support case $b_n=b$.}
  \label{withbound}
\end{figure}

Finally, we would like to study the case in which the Lanczos coefficient has a bounded support. In this case, the continuum limit analysis at leading order is failed to capture the leading time dependence of K-entropy (but the result for K-complexity is still valid) \cite{Barbon:2019wsy} . Hence, our result for $\Delta S_K$ in sec.\ref{continuum} breaks down and the bound proposed in (\ref{boundsk}) looks problematic in this case. However, our numerical results still support that $\Delta S_K\rightarrow \mm{const}$ in the long time limit and the bound (\ref{boundsk}) continues hold.  For example, consider the simplest case $b_n=b$ which has $\eta\simeq 0.729302$. The wave function is solved in terms of Bessel functions
\be \varphi_n(t)=J_n(2b t)+J_{n+2}(2bt) \,.\ee
The numerical results for entropy growth are shown in Fig.\ref{withbound}. It is immediately seen that up to subleading oscillations, $\Delta S_K$ approaches to a constant while the entropy growth rate still decreases in a power law $\partial_t S_K\sim b_1\Delta K/K\sim 1/t$.

Combining all the above results together, we conclude that during irreversible operator growth, the K-entropy growth can be roughly classified into two regimes: 1) the initial regime, in which the entropy growth rate nearly saturates the dispersion bound $\partial_t S_K\simeq 2b_1 \Delta S_K$; 2) the late time regime, in which the growth rate can be well described by the new bound (\ref{boundsk}) $\partial_t S_K\sim 2b_1\eta\Delta K/K$. For physically interesting cases, such as chaotic systems and integrable models, this describes the growth of K-entropy very well during the Heisenberg evolution.

\section{Conclusion}

In this paper, we examine the upper bound on the growth of general A-complexity defined in the Krylov space, including K-complexity and K-entropy, during the Heisenberg evolution of operators. We first prove a dispersion bound (\ref{boundgene}), generalising the result of K-complexity \cite{Hornedal:2022pkc} to general A-complexity. Our new contributions are: 1) We show that saturation of the dispersion bound on A-complexity is possible if and only if the A-complexity operator $\mc{A}$ is linearly related to the K-complexity operator $\mc{K}$: $\mc{A}=\alpha \mc{K}+\beta$, where $\alpha\,,\beta$ are c-number constants (independent of $n$ but may depend on time). However, since saturation of the bound for K-complexity is equivalent to closure of the complexity algebra \cite{Hornedal:2022pkc}, this implies that the complexity algebra constrains the growth of general A-complexity in a subtle way. 2) Though there are only a few cases in which the complexity algebra is closed, the K-complexity grows sufficiently fast at long times for general irreversible process, with the rate of change close to the dispersion bound $\partial_t K\sim 2b_1\Delta K$.

However, the situation for K-entropy is quite different. Except chaotic systems, the entropy growth rate generally decreases in a power law to leading order at long times $\partial_t S_K\sim 1/t^\gamma\,,0<\gamma\leq 1$ ($\gamma=0$ for chaotic systems ) whereas the variance $\Delta S_K$ approaches to a constant in the long time limit. Thus generally the dispersion bound on K-entropy is too loose to be saturated. However, it turns out that the dispersion bound of K-complexity leads to a tighter bound on the growth of K-entropy (\ref{boundsk}) in the long time limit because of a universal logarithmic relation between K-complexity and K-entropy for irreversible process \cite{Fan:2022xaa}. In particular, for physically interesting cases such as chaotic systems and integrable models, the K-entropy growth rate at long times is very close to this new bound.

\section*{Acknowledgments}

Z.Y. Fan was supported in part by the National Natural Science Foundations of China with Grant No. 11873025.

\end{document}